\documentclass{article}
\usepackage{amsmath}

\newtheorem{theorem}{Theorem}
\newtheorem{acknowledgement}[theorem]{Acknowledgement}

\input{tcilatex}
\begin{document}

\title{Cosmology: the search for twenty-four\ (or more) functions}
\author{John D. Barrow \\
%EndAName
DAMTP, University of Cambridge\\
Wilberforce Rd., Cambridge CB3 0WA\\
UK}
\date{}
\maketitle

\begin{abstract}
We enumerate the 4(1+F)+2S independent arbitrary functions of space required
to describe a general relativistic cosmology containing an arbitrary number
of non-interacting fluid (F) and scalar fields (S). Results are also given
for arbitrary space dimension and for higher-order gravity theories, where
the number increases to 16+4F+2S. Both counts are subject to assumptions
about whether the dark energy is a cosmological constant. A more detailed
analysis is provided when global homogeneity is assumed and the functions
become constants. This situation is also studied in the case where the flat
and open universes have compact spatial topologies. This changes the
relative generalities significantly and places new constraints on the types
of expansion anisotropy that are permitted. The most general compact
homogeneous universes containing Friedmann models are spatially flat and
described by 8+4F+2S constants. Comparisons are made with the simple
6-parameter lambda-CDM model and physical interpretations provided for the
parameters needed to describe the most general cosmological models.
\end{abstract}

\section{ Introduction}

There have been several attempts to reduce the description of the
astronomical universe to the determination of a small number of measurable
parameters. Typically, these will be the free parameters of a well defined
cosmological model that uses the smallest number of constants that can
provide a best fit to the available observational evidence. Specific
examples are the popular characterisations of cosmology as a search for
'nine numbers' \cite{rrob}, 'six numbers' \cite{rees}, or the six-parameter
minimal $\Lambda CDM$ model used to fit the WMAP \cite{wmap} and Planck data
sets \cite{Planck}. In all these, and other, cases of simple parameter
counting there are usually many simplifying assumptions that amount to
ignoring other parameters or setting them to zero; for example, by assuming
a flat Friedmann background universe or a power-law variation of density
inhomogeneity in order to reduce the parameter count and any associated
degeneracies. The assumption of a power-law spectrum for inhomogeneities
will reduce a spatial function to two constants, while the assumption that
the universe is described by a Friedmann metric plus small inhomogeneous
perturbations both reduces the number of metric unknowns and converts
functions into constant parameters. In this paper we are going provide some
context for the common minimal parameter counts cited above by determining
the total number of spatial functions that are needed to prescribe the
structure of the universe if it is assumed to contain a finite number of
simple matter fields. We are not counting fundamental constants of physics,
like the Newtonian gravitation constant, the coupling constants defining
quadratic lagrangian extensions of general relativistic gravity, or the 19
free parameters that define the behaviours of the 61 elementary particles in
the standard 3-generation $U(1)\times SU(2)\times SU(3)$ model of particle
physics. However, there is some ambiguity in the status in some quantities.
For example, whether the dark energy is equivalent to a true cosmological
constant (a fundamental constant), or to some effective fluid or scalar
field, or some other emergent effect \cite{shaw}? Some fundamental physics
parameters, like neutrino masses, particle lifetimes, or axion phases, can
also play a part in determining cosmological densities but that is a
secondary use of the cosmological observable. Here we will take an
elementary approach that counts the number of arbitrary functions needed to
specify the general solution of the Einstein equations (and its
generalisations).This will give a minimalist characterisation that can be
augmented by adding any number of additional fields in a straightforward
way. We will also consider the count in higher-order gravity theories as
well as for general relativistic cosmologies. We enumerate the situation in
spatially homogeneous universes in detail so as to highlight the significant
impact of their spatial topology on evaluations of their relative generality.

\section{Simple Function Counting}

The cosmological problem can be formulated in general relativity using a
metric in a general synchronous reference system \cite{LL}. Assume that
there are $F$ matter fields which are non-interacting and each behaves as a
perfect fluid with some equation of state $p_{i}(\rho _{i})$, $i=1,...F$.
They will each have a normalised 4-velocity field, $(u_{a})_{i}$, $%
a=0,1,2,3. $ These will in general be different and non-comoving. Thus each
matter field is defined on a spacelike surface of constant time by $4$
arbitrary functions of three spatial variables, $x^{\alpha }$ since the $%
u_{0}$ components are determined by the normalisations $(u_{a}u^{a})_{i}=1$.
This means that the initial data for the $F$ non-interacting fluids are
specified by $4F$ functions of three spatial variables. If we were in an $N$%
-dimensional space then each fluid would require $N+1$ functions of $N$
spatial variables and $F$ fluids would require $(N+1)F$ such functions to
describe them in general.

The 3-d metric requires the specification of $6$ $g_{\alpha \beta \text{ \ }%
} $and $6$ $\dot{g}_{\alpha \beta }$ for the symmetric spatial $3\times 3$
metric in the synchronous system but these may be reduced by using the $4$
coordinate covariances of the theory and a further $4$ can be eliminated by
using the $4$ constraint equations of general relativity. This leaves $4$
independently arbitrary functions of three spatial variables \cite{LL} which
is just twice the number of degrees of freedom of the gravitational spin-2
field. The general transformation between synchronous coordinate systems
maintains this number of functions \cite{LL}. This is the number required to
specify the general vacuum solution of the Einstein equations in a $3$%
-dimensional space. In an $N$-dimensional space we would require $N(N+1)$
functions of $N$ spatial variables to specify the initial data for $%
g_{\alpha \beta \text{ \ }}$and $\dot{g}_{\alpha \beta }$. This could be
reduced by $N+1$ coordinate covariances and $N+1$ constraints to leave $%
(N-2)(N+1)$ independent arbitrary functions of $N$ variables \cite{JDB}.
This even number is equal to twice the number of degrees of freedom of the
gravitational spin-2 field in $N+1$ dimensional spacetime .

When we combine these counts we see that the general solution in the
synchronous system for a general relativistic cosmological model containing $%
F$ fluids requires the specification of $(N-2)(N+1)+F(N+1)=(N+1)(N+F-2)$
independent functions of $N$ spatial variables. If there are also $S$
non-interacting scalar fields, $\phi _{j}$, $j=1,..,S,$ present with self
interaction potentials $V(\phi _{j})$ then two further spatial functions are
required ($\phi _{j}$ and $\dot{\phi}_{j}$) to specify each scalar field and
the total becomes $(N+1)(N+F-2)+2S$. For the observationally relevant case
of $N=3$, this reduces to $4(F+1)+2S$ spatial functions.

For example, if we assume a simple realistic scenario in which the universe
contains separate baryonic, cold dark matter, photon, neutrino and dark
energy fluids, all with separate non-comoving velocity fields, but no scalar
fields, then $F=$ $5$ and our cosmology needs $24$ spatial functions in the
general case. If the dark energy is not a fluid, but a cosmological constant
with constant density and $u_{i}=\delta _{i}^{0},$ then the dark energy
'fluid' description reduces to the specification of a single constant, $\rho
_{DE}=\Lambda /8\pi G$, rather than $4$ functions and reduces the total to $%
21$ independent spatial functions. However, if the cosmological constant is
an evolving scalar field then we would have $F=4$ and $S=1$, and now $22$
spatial functions are required. Examples of full function asymptotic
solutions were found for perturbations around de Sitter space-time by
Starobinsky \cite{star}, the approach to 'sudden' finite-time singularities 
\cite{sudd} by Barrow, Cotsakis and Tsokaros \cite{BC}, and near
quasi-isotropic singularities with $p>\rho $ 'fluids' by Heinzle and Sandin 
\cite{heinz}.

These function counts of $21$-$24$ should be regarded as lower bounds. They
do not include the possibility of a cosmological magnetic field or some
other unknown matter fields. They also treat all light ($<<1MeV$) neutrinos
as if they are identical (heavy neutrinos can be regarded as CDM if they
provide the largest contribution to the matter density but if they are not
responsible for the dominant dark matter then they should be counted as a
further contribution to $F$). If there are matter fields which are not
simple fluids with $p(\rho ),$for example an imperfect fluid possessing a
bulk viscosity or a gas of free particles with anisotropic pressures, then
additional parameters are required to specify them -- although there can
still be overall constraints -- a trace-free energy-momentum tensor, for
example, in the cases of electric and magnetic fields or Yang-Mills fields
-- and we would just count the number of independent terms in the symmetric
energy-momentum tensor.

In the case of the Planck or WMAP mission data analyses, $6$ constants are
chosen to define the standard (minimal) $\Lambda CDM$ model. For WMAP \cite%
{wmap}, these are the present-day Hubble expansion rate, $H_{0}$, the
densities of baryons and cold dark matter, the optical depth, $\tau $, at a
fixed redshift, and the amplitude and slope of an assumed power-law spectrum
of curvature inhomogeneities on a specified reference length scale. This is
equivalent to including three matter fields (radiation, baryons, cold dark
matter) but the standard $\Lambda CDM$ assumes zero spatial curvature, $k$, 
\textit{ab initio} so a relaxation of this would add a curvature term or a
dark energy field, because when $k\neq 0$ the latter could no longer be
deduced from the other densities and the critical density defined by $H_{0}$%
.The light neutrino densities are assumed to be calculable from the
radiation density using the standard isotropically expanding thermal
history, so there are effectively $F=5$ matter fields (with $k$ set to zero
in the base model) and a metric time derivative determined by $H$). All
deviations from isotropy and homogeneity enter only at the level of
perturbation theory and are characterised by the spectral amplitude and
slope on large scales; the amplitude on small scales ('acoustic peaks' in
the power spectrum) is determined from that on large scales by an $e^{-2\tau
}$ damping factor determined by the optical depth parameter $\tau $. The
Planck parameter choice is equivalent \cite{Planck}.

Although a general solution of the Einstein equations requires the full
complement of arbitrary functions, different parts of the general solution
space can have behaviours of quite different complexity. For example, when $%
N\leq 9$ there are homogeneous vacuum universes which are dynamically
chaotic but the chaotic behaviour disappears when $N\geq 10$\ even though
the number of arbitrary constants remains maximal for each $N$ \cite{chaos}.

\section{\protect\bigskip More General Gravity Theories}

There has been considerable interest in trying to explain the dark energy as
a feature of a higher-order gravitational theory that extends the lagrangian
of general relativity in a non-linear fashion \cite{BO, far, ferr, clif}.
This offers the possibility of introducing a lagrangian that is a function
of $L=f(R,R_{ab}R^{ab})$ of the scalar curvature $R$ and/or the Ricci scalar 
$R_{ab}R^{ab}$ in anisotropic models, with the property that it contributes
a slowly varying dark energy-like behaviour at late times without the need
to specify an explicit cosmological constant. However, these higher-order
lagrangian theories (excluding the Lovelock lagrangians in which the
variation of the higher-order terms contribute pure divergences \cite{Love})
all have 4th-order field equations in 3-dimensional space when $f$ $\neq
A+BR,$ with $A,B$ constants. This means that the initial data set for such
theories is considerably enlarged because we must specify $\ddot{g}_{\alpha
\beta \text{ }}$ and $\dddot{g}_{\alpha \beta }$ in addition to $g_{\alpha
\beta \text{ \ }}$and $\dot{g}_{\alpha \beta }$. In $N$ dimensions, this
results in a further $N(N+1)$ functions of $N$ variables and so a general
cosmological model with $F$ fluids and $S$ scalar fields requires a
specification of $2(N^{2}-1)+F(N+1)+2S=(N+1)(F+2N-2)$ $+2S$ independent
arbitrary spatial functions. For $N=3,$this is $16+4F+2S$. General
relativity with $4$ matter fields plus a cosmological constant requires $20$
spatial functions plus one constant, in general, whereas a higher-order
gravity theory with $4$ matter fields and no scalar fields (and no
cosmological constant because it should presumably emerge from the metric
behaviour) requires the specification of $32$ spatial functions in general.

\section{Reducing Functions to Constants}

The commonest simplification used to reduce the size of the cosmological
characterisation problem is to turn the spatial functions into constants.
This simplification will be an exact if the universe is assumed to be
spatially homogeneous. The set of possible spatially homogeneous and
isotropic universes with natural topology is based upon the classification
of homogeneous 3-spaces created by Bianchi \cite{Bi, taub,mac, wain}
(together with the exceptional case of Kantowski-Sachs-Kompanyeets-Chernov
with $S^{1}\times S^{2}$ topology \cite{KS},\cite{KC} which we will ignore
here).

The most general Bianchi type universes are those of types $%
VI_{h},VII_{h},VIII$ and $IX$. \ Of these, only types $VII_{h\text{ }}$and $%
IX$, respectively, contain open and closed isotropic Friedmann subcases.
These most general Bianchi types are all defined by $4$ arbitrary constants
in vacuum plus a further $4$ for each non-interacting perfect fluid source.
Therefore, in three-dimensional spaces, the most general spatially
homogeneous universes containing $F$ fluids are defined by $4(1+F)$
arbitrary constants. This suggests that they might be the leading order term
in a linearisation of the general inhomogeneous solution in the homogeneous
limit. However, things might not be so simple. The 4-function space of
solutions to Einstein's models like type $IX$ with compact spaces has a
conical structure at points with Killing vectors and so linearisation about
the points must control an infinite number of spurious linearisations that
are not the leading-order term in any series expansion that converges to a
true solution \cite{mar, BT}.

The Bianchi classification of spatially homogeneous universes derives from
the classification of the group of isometries with three-dimensional
subgroups that act simply transitively on the manifold. Intuitively, these
give cosmological histories that look the same to observers in different
places on the same hypersurface of constant time.

The Bianchi types are subdivided into two classes \cite{EM}: Class A
contains types\emph{\ }$I(1+F),II(2+3F),VI_{0}(3+4F),VII_{0}(3+4F),VIII(4+4F)
$\emph{\ }and $IX(4+4F),$\ while Class B contains types $%
V(1+4F),IV(3+4F),III(3+4F),VI_{-1/9}(4+3F),VI_{h}(4+4F)$\ and $VII_{h}(4+4F).
$ The brackets following each Roman numeral labeling the Bianchi type
geometry contain the number of constants defining the general solution when $%
F$ non-interacting perfect fluids, each with $p>-\rho $, are present, so $F=0
$ defines the vacuum case. For example, Bianchi type I denoted by $I(1+F)$
is defined by one constant in vacuum (when it is the Kasner metric) and one
additional constant for the value of the density when each matter field is
added. We have ignored scalar fields here for simplicity but to restore that
consideration simply add $2S$ inside each pair of brackets. The Euclidean
geometry in the type $I$ case requires $R_{0\alpha }=0$ and so the $3$
non-comoving velocities (and hence any possible vorticity) must be
identically zero. This contains the zero-curvature Friedmann model as the
isotropic (zero parameter) special case. In the next simplest case, of type $%
V$, the general vacuum solution found by Saunders \cite{sau} contains one
parameter, but each additional perfect-fluid adds $4$ parameters because it
requires specification of a density and three non-zero $u_{\alpha }$
components. The spatial geometry is a Lobachevsky space of constant negative
isotropic curvature. The isotropic subcases of type $V$ are the
zero-parameter Milne universe in vacuum and the $F$-parameter open Friedmann
universe containing $F$ fluids.

In practice, one cannot find exact homogeneous general solutions containing
the maximal number of arbitrary constants, although the qualitative
behaviours are fairly well understood, and many explorations of the
observational effects use the simplest Bianchi I or V models (usually
without including non-comoving velocities) because they possess isotropic
3-curvature and add only a simple fast-decaying anisotropy term (one
constant parameter) to the Friedmann equation. The most general anisotropic
metrics which contain isotropic special cases, of types $VII$ and $IX$,
possess both expansion anisotropy (shear) and anisotropic three-curvature.
Their shear falls off more slowly and the observational bounds on it are
much weaker \cite{CH},\cite{dln},\cite{bbn},\cite{bjs},\cite{pep, skew}.

\section{Links to Observables}

The free spatial functions (or constants) specifying inhomogeneous
(homogeneous) metrics have simple physical interpretations. In the most
general cases the $4$ vacuum parameters can be thought of as giving two
shear modes (ie time-derivatives of metric anisotropies) and two parts of
the anisotropic spatial curvature (composed of ratios and products of metric
functions). In the simplest vacuum models of type $I$ and $V$ the
three-curvature is isotropic so there is only one shear parameters. It
describes the allowed metric shear and in the $V$ model a second parameter
is the isotropic three-curvature (which is zero in type $I$). When matter is
added then there is always a single $\rho ($or $p$) for each perfect fluid
and then up to three non-comoving fluid velocity components. If the fluid is
comoving, as in type $I$ only the density parameter is required for each
fluid; in type $V$ there can also be $3$ non-comoving velocities. The
additional parameters control the expansion shear anisotropy, anisotropic
3-curvature. They may all contribute to temperature anisotropy in the CMB
radiation but the observed anisotropy is determined by an integral down the
past null cone over the shear (effectively the shear to Hubble rate ratio at
last scattering of the CMB), rather than the Weyl curvature modes driven by
the curvature anisotropy (which can be oscillatory \cite{wain2}, and so can
be periodically be very small), while the velocities contribute dipole
variations. Thus, it is difficult to extract complete information about all
the anisotropies from observations of the lower multipoles of the CMB alone
in the most general cases \cite{djs, WS, M, Nil}.

At present, the observational focus is upon testing the simplest possible $%
\Lambda CDM$ model, defined by the smallest number of parameters. As
observational sensitivity increases it will become possible to place
specific bounds or make determinations of the full spectrum of defining
functions (or constants). In an inflationary model they can be identified
from the spatial functions defining the asympotic expansion around the de
Sitter metric \cite{star}.

There have also been interesting studies of the observational information
needed to determine the structure of our past null cone rather than
constant-time hypersurfaces in the Universe \cite{ell}, extending earlier
investigations of the links between observables and general metric
expansions by McCrea \cite{mc} and by Kristian and Sachs \cite{ks}. The high
level of isotropy in the visible universe, possibly present as a consequence
of a period of inflation in the early universe \cite{infl}, or special
initial conditions \cite{spec, pep, heinz, pen, swh}, is what allows several
of the defining functions of a generic cosmological model to be ignored on
the grounds that they are too small to be detected with current technology.
An inflationary theory of the chaotic or eternal variety, in which inflation
only ends locally, will lead to some complicated set of defining functions
that exhibit large smooth isotropic regions within a complicated global
structure which is beyond our visual horizon and unobservable (although not
necessarily falsifiable within a particular cosmological model).

\ \ In practice, there is a divide between the complexity of inhomogeneity
in the universe on small and large scales. On large scales there has been
effectively no processing of the primordial spectrum of inhomogeneity by
damping or non-linear evolution. Its description is well approximated by
replacing a smooth function by a power-law defined by 2 constants, as for
the microwave background temperature fluctuation spectrum or the 2-point
correlation function of galaxy clustering. Here, the defining functions may
be replaced by statistical distributions for specific features, like peak or
voids in the density distribution. On small scales, inhomogeneities that
entered the horizon during the radiation era can be damped out by photon
viscosity or diffusion and may leave distortions in the background radiation
spectrum as witness to their earlier existence. The baryon distribution may
provide baryon acoustic oscillations which yield potentially sensitive
information about the baryon density \cite{wmap, Planck}. On smaller scales
that enter the horizon later, where damping and non-linear self-interaction
has occurred, the resulting distributions of luminous and dark matter are
more complicated. However, they are correspondingly more difficult to
predict in detail and numerical simulations of ensembles of models are used
to make predictions down to the limit of reliable resolution. Predicting
their forms also requires a significant extension of the simple, purely
cosmological enumeration of free functions that we have discussed so far.
Detailed physical interactions, 3-d hydrodynamics, turbulence, shocks,
protogalaxy shapes, magnetic fields, and collision orientations, all
introduce additional factors that may increase the parameters on which
observable outcomes depend. The so called bias parameter, equal to the ratio
of luminous matter density to the total density, is in reality a spatial
function that is being used to follow the ratio of two densities because one
(the dark matter) is expected to be far more smoothly distributed than the
other. All these small scale factors combine to determine the output
distribution of the baryonic and non-baryonic density distributions and
their associated velocities.

\section{Topology}

So far, we have assumed that the cosmological models in question take the
natural topology, that is $R^{3}$ for the 3-dimensional flat and negatively
curved spaces and $S^{3}$ for the closed spaces. However, compact topologies
can also be imposed upon flat and open universes and there has been
considerable interest in this possibility and its observational consequences
for optical images of galaxies and the CMB, \cite{ellis, top, top2}. The
classification of compact negatively-curved spaces is a challenging
mathematical problem and when the permitted compact spatial topologies are
imposed on flat and open homogeneous cosmologies it produces a major change
in their relative generalities and the numbers of constants needed to
specify them in general and in the differences between the counts for vacuum
and perfect fluid models.

The most notable consequences of imposing a compact topology on
3-dimensional homogeneous spaces is that the Bianchi universes of types $IV$
and $VI_{h}$ no longer exist and open universes of Bianchi types $V$ and $%
VII_{h}$ must be isotropic with spaces that are quotients of a space of
constant negative curvature, as required by Mostow's Rigidity theorem \cite%
{Ash,fagu,BK1,BK2,K}. The only universes with non-trivial structure that
differs from that of their universal covering spaces are those of Bianchi
types $I,II,III,VI_{0},VII_{0}$ and $VIII$. The numbers of parameters needed
to determine their general cosmological solutions when $F$ non-interacting
fluids are present and the spatial geometry is compact are given by $%
I(10+F),II(6+3F),III(2+N_{m}^{\prime }+F),VI_{0}(4+4F),VII_{0}(8+4F)$ and $%
VIII(4+N_{m}+4F)$, again with $F=0$ giving the vacuum case, as before, and
an addition of $2S$ to each prescription if $S$ scalar fields are included.
Here, $N_{m}$ is the number of moduli degrees of freedom which measures of
the complexity of the allowed topology, with $N_{m}\equiv 6g+2k-6\equiv
N_{m}^{\prime }-2g$, where $g$ is the genus and $k$ is the number of conical
singularities of the underlying orbifold \cite{BK1,BK2}. It can be
arbitrarily large. The rigidity restriction that compact types $V$ and $%
VII_{h}$ must be isotropic means that compactness creates general parameter
dependencies of $V(F)$ and $VII_{h}(F)$ which are the same as those for the
open isotropic Friedmann universe, or the Milne universe in vacuum when $F=0$%
. It is possible that a similar restriction occurs for open compact
inhomogeneous universes and they may only exist when they are isotropic and
homogeneous. However, the situation there is much more complicated and the
theorem of Fischer, Marsden and Moncrief indicates that linearisation
instabilities will also beset attempts to perturb away from homogeneous
spaces because of the coexistence of Killing symmetries and spatial
compactness \cite{BT, mar}.

The resulting classification is shown in Table 1. We see that the
introduction of compact topology for the simplest type $I$ spaces produces a
dramatic increase in relative generality. Indeed, they become the most
general vacuum models by the parameter-counting criterion. An additional $9$
parameters are required to describe the compact type $I$ universe compared
to the case with non-compact Euclidean $R^{3}$ topology. The reason for this
increase is that at any time the compact 3-torus topology requires $3$
identification scales in orthogonal directions to define the torus and $3$
angles to specify the directions of the vectors generating this lattice plus
all their time-derivatives. This gives $12$ parameters, of which $2$ can be
removed using a time translation and a constraint equation, leaving $10$ in
vacuum compared to the $1$ required in the non-compact Kasner vacuum case.

The following general points are worth noting: (i) The imposition of a
compact topology changes the relative generalities of homogeneous
cosmologies; (ii) The compact flat universes are more general than the open
or closed ones; (iii) Type $VIII$ universes, which do not contain Friedmann
special cases but can in principal become arbitrarily close to isotropy are
the most general compact universes. The most general case that contains
isotropic special cases is that of type $VII_{0}$ -- recall that the $%
VII_{h} $ metrics are forced to be isotropic so open Friedmann universes now
become asymptotically stable \cite{BK1} and approach the Milne metric
whereas in the non-compact case they are merely stable and approach a family
of anisotropic vacuum plane waves \cite{late}.

Table 1: \textit{The number of independent arbitrary constants required to
prescribe the general 3-dimensional spatially homogeneous Bianchi type
universes containing }$F$\textit{\ perfect fluid matter sources in cases
with non-compact and compact spatial topologies. The vacuum cases arise when 
}$F=0$\textit{. If }$S$ scalar \textit{fields are also present then each
parameter count increases by }$2S$\textit{.} \textit{The type }$IX$\textit{\
universe does not admit a non-compact geometry and compact universes of
Bianchi types }$IV$\textit{\ and }$VI_{h}$\textit{do not exist. Types }$III$%
\textit{\ and }$VIII$\textit{\ have potentially unlimited topological
complexity and arbitrarily large numbers of defining constants parameters
through the unbounded topological parameters }$N_{m}\equiv 6g+2k-6$\textit{\
and }$N_{m}^{\prime }=N_{m}+2g$\textit{, where }$g$\textit{\ is the genus
and }$k$\textit{\ is the number of conical singularities of the underlying
orbifold.}

\bigskip 
\begin{tabular}{|c|c|c|}
\hline
Cosmological & \multicolumn{2}{|c|}{No. of defining parameters with $F$
non-interacting fluids} \\ \cline{2-3}
Bianchi Type & Non-compact topology & Compact topology \\ \hline
$I$ & $1+F$ & $10+F$ \\ \hline
$II$ & $2+3F$ & $6+3F$ \\ \hline
$VI_{0}$ & $3+4F$ & $4+4F$ \\ \hline
$VII_{0}$ & $3+4F$ & $8+4F$ \\ \hline
$VIII$ & $4+4F$ & $4+N_{m}+4F$ \\ \hline
$IX$ & $-$ & $4+4F$ \\ \hline
$III$ & $3+4F$ & $2+N_{m}^{\prime }+F$ \\ \hline
$IV$ & $3+4F$ & $-$ \\ \hline
$V$ & $1+4F$ & $F$ \\ \hline
$VI_{h}$ & $4+4F$ & $-$ \\ \hline
$VII_{h}$ & $4+4F$ & $F$ \\ \hline
\end{tabular}

\section{\protect\bigskip Inhomogeneity}

The addition of inhomogeneity turns the constants defining the cosmological
problem into functions of three space variables. For example, we are
familiar with the linearised solutions for small density perturbations of a
Friedmann universe with natural topology which produces two functions of
space that control temporally growing and decaying modes. The function of
space in front of the growing mode is typically written as a power-law in
length scale (or wave number) and so has arbitrary amplitude and power index
(both usually assumed to be scale-independent constants to first or second
order) which can fitted to observations. Clearly there is no limit to the
number of parameters that could be introduced to characterise the density
inhomogeneity function by means of a series expansion around the homogeneous
model (and the same could be done for any vortical or gravitational-wave
perturbation modes). Further analysis of the function characterising the
radiation density is seen in the attempts to measure and calculate the
deviation of its statistics from gaussianity \cite{NG} and to reconstruct
the past light-cone structure of the universe \cite{M}. Any different choice
of specific spatial functions to characterise inhomogeneity in densities or
gravitational waves requires some theoretical motivation. What happens in
the inhomogeneous case if open or flat universes are given compact spatial
topologies is not known. As we have just seen, the effects of topology on
the spatially homogeneous anisotropic models was considerable whereas the
effects on the overall evolution of isotropic models (as opposed to multiple
image optics) is insignificant. It is generally just assumed that
realistically inhomogeneous universes with non-positive curvature (or
curvature of varying sign) can be endowed with a compact topology and, if
so, this places no constraints on their dynamics. However, both assumptions
would be untrue for homogeneous universes and would necessarily fail for
inhomogeneous ones in the homogeneous limit. it remains to be determined
what topological constraints arise in the inhomogeneous cases. They could be
weaker because inhomogeneous anisotropies can be local (far small in scale
than the topological identifications) or they could be globally constrained
like homogeneous anisotropies. Newtonian intuitions can be dangerous because
compactification of a Newtonian Euclidean cosmological space seems simple
but if we integrate Poisson's equation over the compact spatial volume we
see that the total mass of matter must be zero \cite{newt}.

\section{Conclusions}

We have provided a simple analysis of the number of independent arbitrary
functions of space required to specify a general cosmological model, with
natural topology, containing a specified number of non-interacting fluids ($%
F $) and scalar fields ($S$) on a spacelike surface of constant time. This
number is equal to $4(1+F)+2S$ in general relativistic cosmologies and
increases to $16+4F+2S$ in higher-order gravity theories with three space
dimensions. Generalisations to universes with $N$ space dimensions were also
found. When the assumption of spatial homogeneity is introduced these
maximal counts remain true for the most general cosmologies but the spatial
functions are replaced by constants. We enumerate these constants for each
of the homogeneous Bianchi type universes containing non-interacting fluids
and scalar fields. When the spaces of the flat and open homogeneous
universes are compactified the classification changes dramatically. Some
homogeneous geometries are no longer permitted and other important cases,
including those of the anisotropic universes containing open Friedmann
universes, are constrained to be exactly isotropic. The hierarchy of
generality changes and the number of constants required to specify the most
general compact topologies increases significantly. For universes containing
isotropic particular cases it is largest for the flat universes of\ type $%
VII_{0},$ where $8+4F+2S$ constants are required, and type $I$, where $%
10+F+2S$ are required, but can be arbitrarily large in the case of type $%
VIII $ because of the unlimited topological complexity. How these
topological constraints change when inhomogeneities are present remains an
open question. These results provide a wider context for the parameter
counts in $\Lambda CDM$ where, for CDM, baryons, radiation and neutrinos $F$
is at least $4,$or $5$ depending on assumptions about the nature of the dark
energy but would necessarily be larger with non-standard topology permitted.

\begin{acknowledgement}
Support from the STFC (UK) and the JTF Oxford-Cambridge Philosophy of
Cosmology program is acknowledged.
\end{acknowledgement}


\bigskip

\begin{thebibliography}{99}
\bibitem{rrob} M. Rowan Robinson, \textit{The Nine Numbers of the Cosmos},
(Oxford UP, Oxford,1999).

\bibitem{rees} M.J. Rees, \textit{Just Six Numbers: the deep forces that
shape the universe, (}Phoenix, London, 2001).

\bibitem{wmap} D.N. Spergel et al (WMAP), Ap. J. Supplt. \textbf{148},175,
(2003).

\bibitem{Planck} P.A.R. Ade et al (Planck Collaboration Paper XVI),
arXiv:1303.5076.

\bibitem{shaw} J.D. Barrow and D. Shaw, Phys. Rev. Lett. \textbf{106},
101302 (2011)

\bibitem{LL} L. \ Landau and E.M. Lifshitz, \textit{The Classical Theory of
Fields}, 4th rev. edn. (Pergamon, Oxford, 1975)

\bibitem{JDB} J.D. Barrow, Gravitation and Hot Big Bang Cosmology, In 
\textit{The Physical Universe: The Interface Between Cosmology, Astrophysics
and Particle Physics}', eds. J.D. Barrow, A. Henriques, M. Lago \& M.
Longair, pp.1-20, (Springer-Verlag, Berlin,1991).

\bibitem{star} A.A. Starobinsky, Sov. Phys. JETP Lett. \textbf{37}, 66 (1983)

\bibitem{sudd} J.D. Barrow, Class. Quant. Grav. \textbf{21}, L79 (2004)

\bibitem{BC} J.D. Barrow, S. Cotsakis and A. Tsokaros, Class. Quant. Grav. 
\textbf{27}, 165017 (2010)

\bibitem{heinz} M. Heinzle and P. Sandin, Comm. Math. Phys. \textbf{313},
385 (2012)

\bibitem{chaos} J. Demaret, M. Henneaux and P. Spindel, Phys. Lett. \textbf{%
164}, 27 (1985)

\bibitem{BO} J.D. Barrow and A.C. Ottewill, J. Phys. A \textbf{16}, 2757
(1983)

\bibitem{far} V. Faraoni and T. Sotiriou, Rev. Mod. Phys. \textbf{82},
451(2010)

\bibitem{ferr} T. Clifton, P. G. Ferreira, A. Padilla, C. Skordis, Phys.
Reports \textbf{513}, 1 (2012)

\bibitem{clif} T. Clifton and J.D. Barrow, Phys. Rev. D \textbf{72},123003
(2005)

\bibitem{Love} D. Lovelock, J. Math. Phys. \textbf{12}, 498 (1971)

\bibitem{Bi} L. Bianchi, Mem. Matematica Fis. d. Soc. Ital. delle Scienza,
Ser. Terza 11, 267 (1898) reprinted in Gen. Rel. Grav. \textbf{33}, 2171
(2001)

\bibitem{taub} A.H. Taub, Ann. Math. \textbf{53}, 472 (1951)

\bibitem{mac} M.A.H. MacCallum, in \textit{General Relativity: An Einstein
Centenary Survey}, eds. S.W. Hawking and W. Israel, (Cambridge UP,
Cambridge, 1979), pp.533-576.

\bibitem{wain} G.F.R. Ellis, S.T.C. Siklos and J. Wainwright, in \textit{%
Dynamical Systems in Cosmology}, eds. J. Wainwright and G.F.R. Ellis
(Cambridge UP, Cambridge, 1997), pp. 11-42.

\bibitem{KS} R. Kantowski and R.K. Sachs, J. Math. Phys. \textbf{7},
443(1966)

\bibitem{KC} A.S. Kompaneets and A.S. Chernov, Sov. Phys. JETP \textbf{20}%
,1303 (1964)

\bibitem{mar} A.E. Fischer, J.E. Marsden \ and V. Moncrief, Ann. Inst. H.
Poincar\'{e} 33,147 (1980)

\bibitem{BT} J.D. Barrow and F.J. Tipler, Phys. Reports \textbf{56}, 371
(1979)

\bibitem{EM} G.F.R. Ellis and M.A.H. MacCallum, Comm. Math. Phys. \textbf{12}%
,108 (1969)

\bibitem{sau} P.T. Saunders, Mon. Not. R. astron. Soc.\textbf{142}, 213
(1969)

\bibitem{CH} C.B. Collins and S.W. Hawking, Mon. Not. R. astron. Soc. 162,
307 (1972)

\bibitem{dln} A. Doroshkevich, V. Lukash and I.D. Novikov, Sov. Phys. JETP
37, 739 (1973)

\bibitem{bbn} J.D. Barrow, Mon. Not. R. astron. Soc. 175, 359 (1976)

\bibitem{djs} J.D. Barrow, R. Juszkiewicz and D.N. Sonoda, Mon. Not. R.
astron. Soc. \textbf{213}, 917 (1985)

\bibitem{pep} J.D. Barrow, Phys. Rev. D \textbf{51}, 3113 (1995)

\bibitem{skew} J.D. Barrow, Phys. Rev. D \textbf{55}, 7451 (1997)

\bibitem{wain2} W.C. Lim, U.S. Nilsson and J. Wainwright, Class. Quant.
Grav. \textbf{18}, 5583 (2001)

\bibitem{WS} W. Stoeger, M. Araujo, T. Gebbie, Ap.J. \textbf{476}, 435 (1997)

\bibitem{M} R. Maartens, Phil. Trans. R. Soc. A \textbf{369}, 5115 (2011)

\bibitem{Nil} U.S. Nilsson, C. Uggla, J. Wainwright and W. C. Lim, Ap. J. 
\textbf{522}, L1 (1999)

\bibitem{ell} G.F.R. Ellis , S.D. Nell, R. Maartens, W.R. Stoeger and A.P.
Whitman, Phys, Rep. \textbf{124}, 315417 (1985)

\bibitem{mc} W.H. McCrea, Zeit. Astrophys. \textbf{9}, 290 (1934) and Zeit.
Astrophys.\textbf{18}, 98 (1939), reprinted in Gen. Rel. Grav. \textbf{30},
315 (1998)

\bibitem{ks} J. Kristian and R.K. Sachs, Ap. J. \textbf{143}, 379 (1966)

\bibitem{infl} A. Guth, Phys. Rev. D \textbf{23}, 347 (1981)

\bibitem{spec} J.D. Barrow, Nature \textbf{272}, 211 (1978)

\bibitem{pen} R. Penrose, \textit{Cycles of Time}, (Bodley Head, London,
2010); P. Tod, arXiv:1309.7248

\bibitem{swh} J. Hartle and S.W. Hawking, Phys. Rev. D \textbf{28}, 2960
(1983)

\bibitem{ellis} G.F.R. Ellis, Gen. Rel. Gravn. \textbf{2}, 7 (1971)

\bibitem{top} R. Aurich, S. Lustig, F. Steiner and H. Then, Class. Quant.
Grav. \textbf{21}, 4901 (2004)

\bibitem{top2} P.A.R. Ade et al (Planck Collaboration Paper XXVI),
arXiv:1303.5086

\bibitem{Ash} A. Ashtekar and J. Samuel, Class. Quant. Grav. \textbf{8},
2191 (1991)

\bibitem{fagu} H.V. Fagundes, Gen. Rel. Gravn. \textbf{24}, 199 (1992)

\bibitem{BK1} J.D. Barrow and H. Kodama, Int. J. \ Mod. Phys. D \textbf{10},
785 (2001)

\bibitem{BK2} J.D. Barrow \ and \ H. Kodama, Class. Quant. Grav.\textbf{18}%
,1753 (2001)

\bibitem{K} H. Kodama, Prog.Theor. Phys. \textbf{107}, 305 (2002)

\bibitem{late} J.D. Barrow and D.H. Sonoda, Phys. Reports \textbf{139}, 1
(1986)

\bibitem{NG} P.A.R. Ade et al (Planck Collaboration Paper XXIV),
arXiv:1303.5084

\bibitem{newt} This follows from Poisson's equation since $0=\int_{V}\nabla
^{2}\Phi dV=4\pi G\int_{V}\rho dV=4\pi GM$, where $V$ is the compact spatial
volume and $\Phi $ is the Newtonian graviational potential.
\end{thebibliography}
\end{document}